\documentclass{iaus}
\usepackage{graphicx}

\title[When do early-type galaxies form?] 
{When do early-type galaxies form?}

\author[R. G. Abraham and the GDDS Team]   
{Roberto G. Abraham$^1$, Patrick J. McCarthy$^2$, Erin Mentuch$^1$, Karl Glazebrook$^3$,
Preethi Nair$^1$, Jean-Ren\'e Gauthier$^1$, Sandra Savaglio$^4$, David Crampton$^5$, Stephanie Juneau$^{5,6}$, Richard Murowinski$^5$, Damien Le Borgne$^1$, R. G. Carlberg$^1$, Inger J{\o}rgensen$^7$, Kathy Roth$^7$, Hsiao-Wen Chen$^8$, Ronald O. Marzke$^9$}

\affiliation{$^1$Department of Astronomy \& Astrophysics, University of Toronto\break
$^2$Observatories of the Carnegie Institution of Washington\break
$^3$Centre for Astrophysics and Supercomputing, Swinburne University of Technology\break
$^4$Max-Planck-Institut f\"ur extraterrestrische Physik\break
$^5$Herzberg Institute of Astrophysics\break
$^6$Steward Observatory, University of Arizona\break
$^7$Gemini Observatory\break
$^8$Dept. of Astronomy \& Astrophysics, University of Chicago\break
$^9$Dept. of Physics and Astronomy, San Francisco State University}

\pubyear{2006}
\volume{235}  
\pagerange{119--126}
\date{?? and in revised form ??}
\jname{Galaxy Evolution across the Hubble Time}
\editors{F.Combes \& J. Palous, eds.}
\begin{document}

\maketitle

\begin{abstract}
We have used the Hubble Space Telescope's Advanced Camera for Surveys to measure the mass density function of morphologically-selected early-type galaxies in the Gemini Deep Deep Survey fields, over the redshift range $0.9<z<1.6$. Our imaging data set
covers four well-separated sight-lines, and is roughly intermediate (in terms of both depth and area) between the GOODS/GEMS imaging data, and the images obtained in the Hubble Deep Field campaigns. Our images contain
144 galaxies with ultra-deep spectroscopy, and they
have been analyzed using a new purpose-written morphological analysis code which
improves the reliability of morphological classifications by adopting a `quasi-petrosian' image thresholding technique. We find that at $z=1$ approximately 70\% of the stars in massive galaxies reside in early-type systems. This fraction is remarkably similar to that seen in the local Universe. However, we detect very rapid evolution in this fraction over the range $1.0<z<1.6$, suggesting that in this epoch the strong color-morphology relationship seen in the nearby Universe is beginning to fall into place.
\end{abstract}

\firstsection 
\section{Introduction}

The era at which a strong correlation between galaxy morphology and stellar content develops is not well constrained. One of the primary goals of the Gemini Deep Deep Survey (GDDS; Abraham et al. 2004) was to use stellar-mass-selected samples to probe galaxy evolution in the critical $1 < z < 2$ range, over which something like half the stellar mass in galaxies is expected to form (based upon integration of the cosmic star-formation history presented in Steidel et al. 1999). The GDDS and other samples (e.g. K20; Cimatti et al. 2004) showed that the total stellar mass density evolves slowly for $z < 1.5$ and that the high-mass end in particular is slowly evolving (Glazebrook et al. 2004; Fontana et al. 2004).  In this short paper we examine the evolution of the stellar mass density as a function of morphological type using deep HST/ACS imaging of the GDDS fields. We outline how a slight refinement (based on adopting a so-called `quasi-petrosian' threshold) to traditional methods allows morphological classification based on central concentration (or Gini coefficent) vs. asymmetry. Using this method,  we are able to derive robust morphological classifications at $1<z<2$, and present the stellar mass function for early-type galaxies over this redshift range. 

\section{Quasi-Petrosian measurements}\label{sec:quasi}

The benefits of determining galaxy properties within a metric size
defined by a circular
Petrosian aperture are well-established (Petrosian 1976;
Blanton et al. 2001), so we will not describe them here
except to note that galaxy properties determined in this way are quite 
robust to the effects of cosmological
dimming and bandshifting. What is less well-known is that a  `Quasi-Petrosian' aperture
can also be constructed using an
algorithm which works for galaxies of arbitrary shape\footnote{We 
learned in Prague that  Ned Taylor at Leiden Observatory has also been
experimenting with quasi-petrosian statistics, which he defines using equations
similar to those given in this section.}. Interestingly,
this algorithm is closely related to the pixel-sorting algorithm 
used to define the Gini coefficient
(Abraham et al. 2003), which itself provides an elegant and robust proxy for
central concentration and which is proving useful
for galaxy classification (Abraham et al. 2003, Lotz et al. 2004)\footnote{Public-domain
software for undertaking morphological measurements
is available at this URL:
{\tt http://odysseus.astro.utoronto.ca/$\sim$abraham/Morpheus/}}.

A `quasi-Petrosian aperture' can be defined in the following manner.
All
pixels in the galaxy  image are
sorted in {\em decreasing} order of flux to construct an array, 
$f_i$, containing the flux in the $i^{\rm th}$ sorted pixel.
This array is then summed over to construct a monotonically
{\em increasing} curve of cumulative flux
values:
\begin{equation} 
\mathcal{F}_i=\displaystyle\sum_{j=1}^i f_j.
\end{equation}
In analogy with the definition of the Petrosian radius for
a circular aperture, a `Petrosian isophote' can be calculated
by determining the pixel index $i$ 
which satisfies the following equation:
\begin{equation}
f_i = q\times \left( \frac{\mathcal{F}_i}{i}\right)
\label{quasiPetroEqn}
\end{equation}
The flux value of the pixel which solves
this equation is used to construct the quasi-petrosian aperture mask. In
fact, once $f_i$ is
known, constructing
the aperture mask is trivial: pixels brighter than $f_i$ are set to 1.0, and pixels
fainter than $f_i$ are set to 0.0. Morphological quantities are computed in
the usual way
after multiplying the original image by the aperture mask.
Measurements obtained through non-circular quasi-Petrosian apertures
share many of the salutary properties of photometric measurements obtained
through circular Petrosian apertures.  The free parameter $q$ in the equation is
fixed for all galaxies in the sample, and ideally all groups using this approach
would adopt the same value of $q$ to allow easy inter-comparison of measurements. A little experimentation
has shown us that $q=0.2$ yields excellent results, and we suggest that this
be adopted as a sensible value. Finally, we note that because the data are discrete,
Equation~\ref{quasiPetroEqn} will rarely be solved exactly, but
a perfectly adequate approximation can be obtained by simply
noting the index of the first zero crossing of 
$f_i - q\times \left( \frac{\mathcal{F}_i}{i}\right)$. 

\begin{figure*}[htbp]
\begin{center}
\includegraphics[width=5.0in]{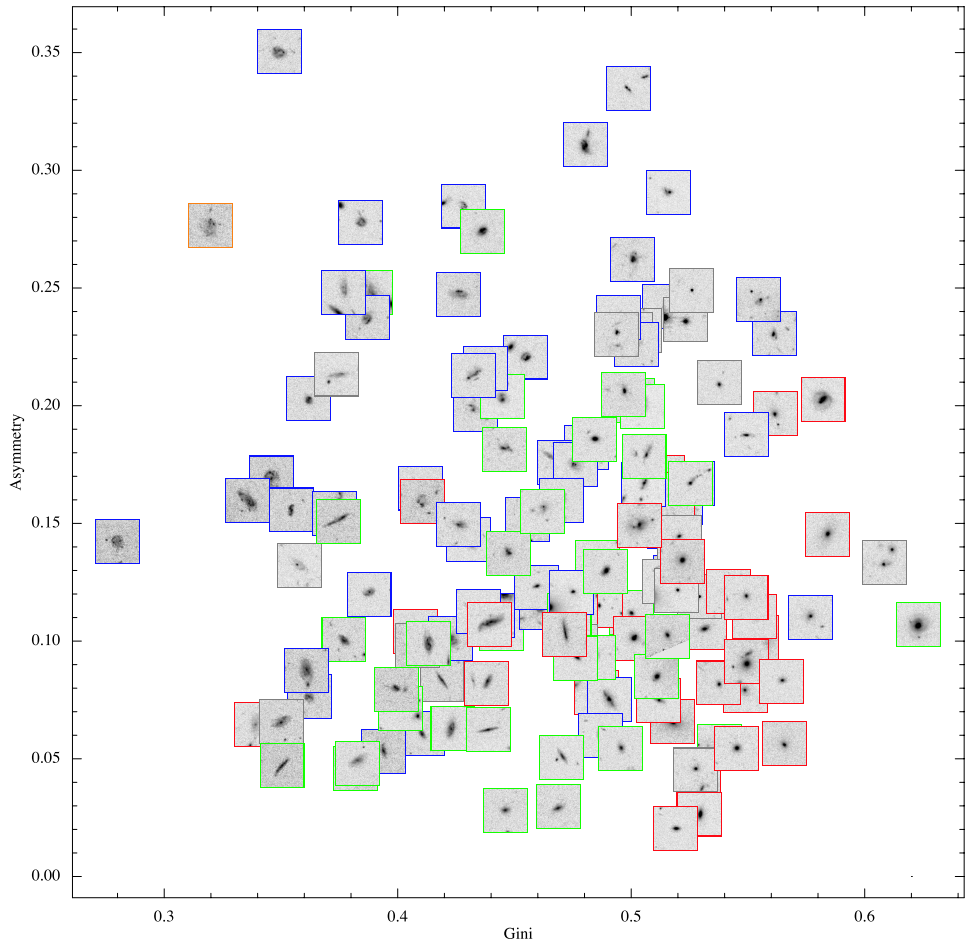} 
\caption{\label{fig:ga}
Asymmetry vs. Gini for galaxies in the Gemini Deep Deep Survey.
Individual galaxies are shown as four arcsec `postage stamps', whose borders
are colored according to spectral classification. Red borders
correspond to galaxies dominated by older stellar populations.
}
\end{center}
\end{figure*}

\section{The evolving mass-density function of Early-type galaxies}

In Figure~1
we show quasi-Petrosian aperture-based measurements of 
Asymmetry vs. Gini coefficent for all
galaxies in with $I<23.5$ mag in our GDDS ACS images
(Abraham et al. 2006). Inspection of this plot (ideally in electronic
form so zooming-in on the small postage stamp images is possible!)
shows that this two-dimensional classification plane does a
good job of distinguishing early-type galaxies from all other
galaxy types. Interestingly, a comparison of the morphological mix with
the spectral classifications of the galaxies shows that surprisingly little 
ambiguity exists in connecting spectral classes
to morphological classes for quiescent systems. This can be
seen in the electronic
version of this paper, which shows this figure using colored borders on
the galaxy images to denote spectroscopic classification.

\begin{figure*}[htbp]
\begin{center}
\includegraphics[width= 4.9in]{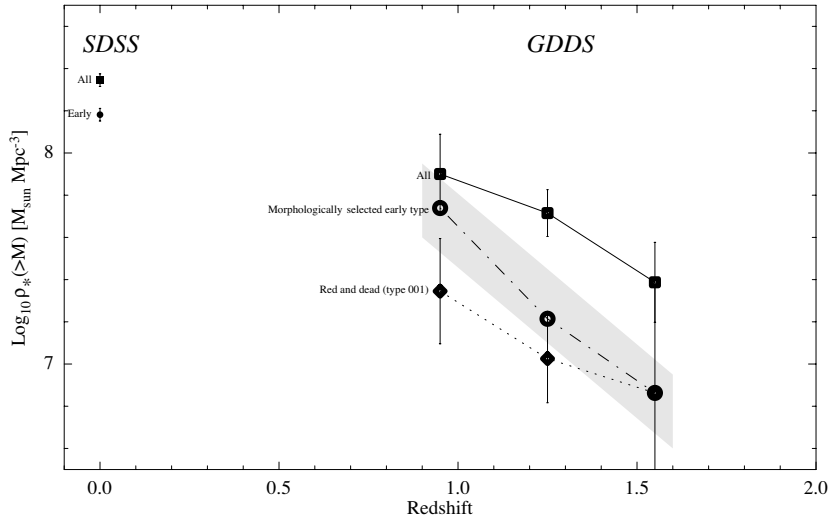} 
\caption{\label{fig:001}
Mass density functions for subsets of massive galaxies, defined as having stellar masses $>10^{10.5} M_\odot$. Plot symbols denote the following subsets:  spectroscopically quiescent `Red and dead' galaxies (diamonds); morphologically selected early-type galaxies (circles); and all galaxies (squares). The gray band corresponds to the approximate uncertainty in the early-type measurment. The local stellar mass densities for massive galaxies, taken from the analysis of SDSS observations given by Bell et al. (2004) and converted to our IMF and mass cut, are also shown.}
\end{center}
\end{figure*}

Figure~2 shows the evolving stellar mass density function for massive 
($\log_{10}(M/M_\odot)>10.5$) 
galaxies in our ACS images, computed using the formalism described in the previous section.
If we restrict consideration to {\em massive galaxies only}, at $z=1$, 
about 70\% of the stellar mass  in these systems lives in early-type galaxies. This is
essentially identical to the local value (Abraham et al. 2006). Interestingly, we see 
strong evolution in this fraction over the range $1<z<1.6$, as the mass densities in spheroids
evolve steeply over this redshift range. These results strongly suggest that the core formation epoch for {\em massive} spheroids is drawing to a close in this interval. Signatures of major mergers appear to have faded by $z \sim 1.5$ or higher, and subsequent mass evolution at $z < 1$ apparently does not significantly disturb the spectral-morphology correlation. 

\begin{acknowledgments}
We thank Francoise Combes for her indulgence in accepting this manuscript even
though it was late in coming, and extend our congratulations to the
local organizers of the IAU symposium in
Prague for hosting an interesting meeting.
\end{acknowledgments}

\end{document}